\documentclass{iopart}
\usepackage{graphicx}

\begin{document}
\title[BEC diffuse reflection]{Diffuse reflection of a Bose-Einstein
condensate from a rough evanescent wave mirror}
\author{H{\'e}l{\`e}ne Perrin$^1$, Yves Colombe$^1$,
Brigitte Mercier$^1$, Vincent Lorent$^1$ and
Carsten Henkel$^2$}
\address{$^1$\ Laboratoire de Physique des Lasers, Universit{\'e} 
Paris 13\\
99 avenue Jean-Baptiste Cl{\'e}ment, 93430 Villetaneuse, France}
\vspace*{0.5ex}
\address{$^2$\ Institut f\"{u}r Physik, Universit\"{a}t Potsdam\\
Am Neuen Palais 10, 14469 Potsdam, Germany}
\ead{henkel@uni-potsdam.de}
\begin{abstract}
We present experimental results showing the diffuse reflection of a 
Bose-Einstein condensate from a rough mirror, consisting of a 
dielectric substrate supporting a blue-detuned evanescent wave. The 
scattering is anisotropic, more pronounced in the direction of the 
surface propagation of the evanescent wave. These results agree very 
well with theoretical predictions.
\end{abstract}
%
%
\submitto{\jpb}

\section{Introduction}

The study of the interactions between ultra cold atoms and surfaces 
is of major interest in the context of Bose-Einstein condensation on 
microchips~\cite{Reichel01b,Folman02}. One motivation is to understand
the limitations on integrated matter wave devices due to imperfect
surface fabrication or finite temperature.  For example, it has been
shown that the
quality of the wires used in microfabricated chips is directly 
linked to the fragmentation effects observed in Bose-Einstein 
condensates (BECs) trapped near a 
metallic wire \cite{Zimmermann02a}. Moreover, the thermal 
fluctuations of the current in a metallic surface induce spin 
flip losses in an atomic cloud when the distance to the surface is 
smaller than $10~\mu$m typically \cite{Jones03}.

Dielectric surfaces and evanescent waves have also been explored 
for producing strong confinement. They have the 
advantage of a strong suppression of the spin flip loss mechanism 
compared to metallic structures
\cite{Henkel99c}. With such a system, one can realize mirrors 
\cite{Balykin88}, diffraction gratings \cite{Dalibard96}, 2D traps 
\cite{Ovchinnikov91} or waveguides \cite{Prentiss00}. Experiments 
involving ultra cold atoms from a BEC at the 
vicinity of a dielectric surface have recently made significant 
progress, leading for 
instance to the realization of a two dimensional BEC \cite{Grimm04a},
to the study of atom-surface reflection in the quantum regime
\cite{Pasquini04},
and to sensitive measurements of adsorbate-induced surface 
polarization 
\cite{Cornell04a} and of the Van der Waals/Casimir-Polder surface 
interaction \cite{Vuletic04}.

In this paper, we present experimental results and a related 
theoretical analysis of Bose condensed Rubidium atoms interacting 
with the 
light field of an evanescent wave above a dielectric slab. The 
evanescent wave is detuned to the blue of an atomic transition line 
and provides a mirror for a BEC that is released
from a trap and falls freely in the gravity field of the earth.
After the bounce off the mirror, we observe a strong scattering of the
atomic cloud (diffuse mirror reflection) that is due to the roughness
of the slab surface where the evanescent wave is
formed~\cite{Landragin96b}. In our case, the phase front of the reflected 
matter waves is significantly distorted because the effective 
corrugation of the mirror is comparable to $\lambda_{\rm dB}/4 \pi 
\cos\theta$ where $\lambda_{\rm dB}$ is the incident de Broglie 
wavelength and $\theta$ the angle of incidence. This is similar
to early experiments with evanescent waves \cite{Landragin96b}
and with magnetic mirrors \cite{HindsHughes}. 
We mention that later experiments achived a significantly reduced 
diffuse reflection (Arnold \etal \cite{Arnold02}) and were even able to
distinguish a specularly reflected matter wave (Savalli \etal
\cite{Savalli2002}). The key result of our experiment is that
we can quantitively confirm the theoretical analysis developed
by Henkel \etal \cite{Henkel97a}, combining independent measurements 
of the dielectric surface and the bouncing atoms.

The paper starts with a presentation of
the experiment and an analysis of the experimental results,
following Ref.\cite{Perrin05a}. We then outline an improved 
theoretical analysis based on Ref.\cite{Henkel97a} and discuss the
momentum distribution of the reflected atoms, in particular its
diffuse spread and its isotropy.

\section{Setup}

\begin{figure}[t]
\begin{center}
\includegraphics[height=20mm]{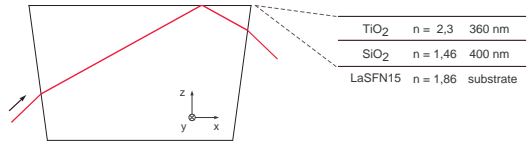}
\end{center}
\caption{Dielectric prism supporting the evanescent wave. The surface 
is coated by two layers of successively low and high refraction index 
to realize a wave guide and enhance the evanescent field. For each 
incident polarization, TE or TM, coupling is resonant for a given 
incident angle. The experiments were performed with TE polarization. 
One denotes $x$ as the propagation axis of the evanescent wave along 
the surface, $y$ as the other horizontal axis and $z$ as the vertical 
one.}
\label{prism}
\end{figure}

The evanescent wave is produced by total internal reflection of a 
Gaussian laser beam at the surface of a dielectric prism. 
As shown in Fig.\ref{prism}, the surface of the prism is coated by 
two 
dielectric layers, a TiO$_2$ layer on top of a SiO$_2$ spacer
layer. This coating forms an optical waveguide that resonantly 
enhances the 
evanescent field above the top layer~\cite{Kaiser94}; we have 
designed 
this configuration for the study of two-dimensional atom
traps~\cite{Perrin03}. The incident angle of the laser beam is fixed 
by the resonance condition for a waveguide mode; for the transverse 
electric (or s) polarization we 
use, the incident angle is $\theta_i = 46.1^{\circ}$ (at the 
TiO$_2$/%
vacuum interface, index $n_{\rm TiO_2} = 1.86$).  The resulting
exponential decay length of the light field is $\kappa^{-1}=93.8$~nm,
and $I = I_0 \, e^{-2 \kappa z}$ is the light intensity.

The mirror light is produced by a laser diode of power 40~mW detuned 
1.5~GHz above the atomic D2 line ($\lambda = 780\,{\rm nm}$ or
$1/\lambda = k_{L}/2\pi = 12\,820\,{\rm cm}^{-1}$). 
The Gaussian beam is elliptical and produces on the surface a spot
with $1/\sqrt{e}$ waist diameters of 
$220~\mu$m along $x$ and $85~\mu$m along $y$ (see coordinate axes in 
Fig.\ref{prism}). A measurement of the reflection 
threshold for the atom beam, taking into account the van der Waals
attraction toward the surface~(Landragin \textit{et al.} in 
Ref.\cite{Balykin88}), gives access to the light intensity at
the surface in the spot center: $I_{0} = 210\,{\rm W/cm}^2$. 
This value is lower than expected from the design of the
dielectric coating; we attribute this to the losses due to
the roughness of the deposited TiO$_2$ layer (see Figure~\ref{AFMpicture} 
below and the discussion there).

\section{Atom bounce}

\subsection{Data}

The experiment proceeds as follows: approximately $10^8$ atoms are
confined in the hyperfine ground state $F=2, m_F=2$ in a Ioffe
Pritchard (IP) type magnetic trap, 3.6~mm above the evanescent mirror
\cite{Perrin03}.  The magnetic trap is cigar shaped, $x$ being its
long axis.  Oscillation frequencies are, respectively, $\omega_x/2 \pi
= 21\,{\rm Hz}$ and $\omega_{\perp}/2 \pi = 220\,{\rm Hz}$ in the
radial directions ($y$ and $z$).  The atoms are evaporatively cooled
to below the
condensation threshold and about $N = 3\times10^5$ atoms are released
at $t = 0$ by switching off the magnetic trapping fields.  These atoms
reach the mirror after free fall at $t_{\mbox{\scriptsize reb}}=27$~ms
and bounce on it
with a velocity $v_{i} = 265\,{\rm mm/s}$ 
(normal incidence $\theta = 0$, de Broglie
wavelength $\lambda_{\rm dB} = 2\pi \hbar / m v_{i} = 17.3\,{\rm
nm}$).
Around the bouncing time $t_{\mbox{\scriptsize reb}}$,
the mirror laser is switched on for $\Delta t = 2.2$~ms. Limiting 
this time window $\Delta t$ prevents near-resonant photon scattering 
during free fall or after reflection. 

\begin{figure}[t]
\begin{center}
\includegraphics[width=70mm]{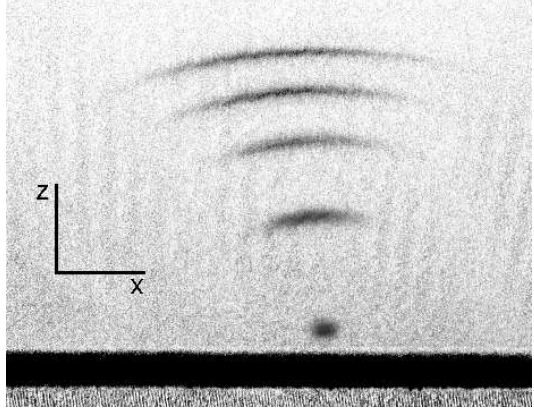}
\end{center}
\caption{Absorption imaging pictures of a bouncing BEC with $3\times 
10^5$ atoms for different times of flight after reflection: 2~ms, 
7~ms, 12~ms, 17~ms and 22~ms.  The pictures, taken with a 200~$\mu$s
long, resonant pulse, are merely superimposed.
The wide black line in the bottom is due to the prism surface, 
slightly tilted from the imaging axis. Picture dimensions are 
5.7~mm $\times$ 4.4~mm.}
\label{bounce}
\end{figure}

The atoms are detected by 
absorption imaging either before or after reflection. During free 
fall, the cloud expands along the radial directions because potential 
and interaction energy is released, but its width along $x$ remains 
nearly constant. The analysis of pictures taken 
before reflection gives access to the following parameters: fraction 
of condensed atoms $N_0/N=0.4$, kinetic temperature of thermal cloud 
$T=285$~nK, initial Thomas-Fermi size along $x$ of the condensed 
fraction
$R_x=90~\mu$m and 
Thomas-Fermi velocity width along $z$: $V_{\perp}=5.96\,{\rm mm/s}$. 
The condensate velocity width along $x$ is very small, thus non 
directly measurable. However, it can be inferred from the know\-ledge 
of $V_{\perp}$ and the oscillation frequencies in the magnetic trap,
using the solution for an expanding BEC \cite{Castin96}; we get
$V_x = \frac{\pi}{2} \, \frac{\omega_x}{\omega_{\perp}} V_{\perp} = 
0.89\,{\rm mm/s}$. The observation of the center of 
mass motion during free fall permits us to calibrate the pixel size 
knowing gravity's acceleration and to infer the initial position 
and velocity of the cloud. The magnetic field switching process 
communicates a small acceleration to the atoms along $x$, resulting 
in a horizontal velocity $v_x = -30.7\,{\rm mm/s}$ (see 
Figure~\ref{bounce}).

After reflection, the absorption images change dramatically 
(figure~\ref{bounce}). The atoms occupy the surface of a scattering 
sphere,
hence an elastic, but strongly diffuse scattering occurs.
For $t > t_{\rm reb}$, the cloud width 
along $x$ increases from its initial value due to an additional 
velocity spread $\sigma_{v_x}$. The 
velocity Gaussian radius at $1/\sqrt{e}$ deduced from the pictures is 
$39.4\,{\rm mm/s}$. Taking into account the initial velocity width 
before reflection, the spread due to diffuse reflection is 
$\sigma_{v_x} = 39\,{\rm mm/s}$, that is 
$6.6\pm0.2~v_{\mbox{\scriptsize 
rec}}$ where $v_{\mbox{\scriptsize rec}} = \hbar k_{L} / m 
= 5.89\,{\rm mm/s}$ is the recoil velocity for Rb. This corresponds 
to 
an angular (rms) spread $\Delta \theta \approx 8.4^\circ$.

The effect of diffuse reflection along $y$ is more subtle to analyze, 
as this axis is aligned with the direction of observation. However, 
it is possible to extract information about $\sigma_{v_y}$ from the 
picture. If for instance the scattering were totally isotropic, with 
$\sigma_{v_y} = \sigma_{v_x}$, the atomic cloud should extend 
asymmetrically towards $-z$ at a given position $x$, as the 
projection of a spherical shell onto a plane extends towards the 
inner part of 
the circle (see figure~\ref{profiles}). If on the contrary the 
scattering would take place only 
along $x$, the cloud width along $z$ at a given position $x$ should 
be very small, with a symmetric shape. 

\subsection{Simulation}

To get some insight into what 
happens along $y$, we performed a numerical simulation of the atomic 
reflection. The simulation calculates $N = 3\times 10^5$ individual 
classical atomic trajectories. The initial positions and velocities 
are chosen to mimic the experimentally measured parameters: 40\% of 
the 
atoms are ``condensed'' and are described by the initial 3D 
Thomas-Fermi
velocity and position distribution. (We neglect the 
position spread along $y$ and $z$ because its contribution 
to the cloud size after a few ms of time of flight is very small.)
The remaining 60\% of the atoms are distributed according to gaussian
profiles for velocity and position, with widths inferred from the 
knowledge 
of temperature and trap parameters. Position and velocity
of the cloud centre are fixed to the experimental values as well.
\begin{figure}[t]
\begin{center}
\includegraphics[width=70mm]{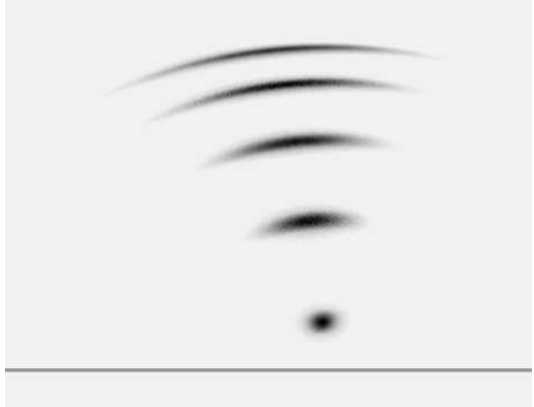}
\end{center}
\caption{Simulation of a bouncing BEC with $3\times 10^5$ atoms for 
the same times of flight than the experimental ones, 
figure~\ref{bounce}. For this series, the velocity spread was chosen 
to be $\sigma_{v_y} = \sigma_{v_x} / 2 = 19.5\,{\rm mm/s}$. The 
position of the mirror surface is marked by a grey line.}
\label{simul}
\end{figure}
The mirror is modelled as an instantaneous diffuse reflector.
This assumption is reasonable as the typical time spent in 
the evanescent wave is small, $1 /\kappa v_i = 0.35~\mu$s. 
After reflection, the atomic velocity is modified to 
describe both specular reflection (inversion of vertical velocity) 
and scattering. A random horizontal velocity is added to the 
reflected velocity with a gaussian distribution. We take a 
$1/\sqrt{e}$ radius $\sigma_{v_{x}} = 39\,{\rm mm/s}$, as measured
experimentally, and run simulations with varying $\sigma_{v_{y}}$.
The $z$ component of the velocity is adjusted in order to preserve 
kinetic energy (the scattering process is elastic, total 
energy is conserved). 
The simulation also takes into account spontaneous emission.
For our parameters, the atom spontaneously emits on average 0.13 photons 
per bounce~\cite{note_spont}. We randomly draw the number of photons 
from a Poisson distribution and add a recoil of $1\,v_{\rm rec}$ in a random 
direction in velocity space for each emission event.
After calculation of all atomic 
trajectories, the atomic density profile is integrated along $y$ as 
in the experimental pictures. We finally apply a Gaussian blur 
filter (width $\sigma_{\mbox{\scriptsize res}} = 9\,\mu$m along $x$ 
and $20~\mu$m along $z$) to mimic the finite resolution of the 
experimental imaging system that we calibrated independently.

\subsection{Anisotropic scattering}

\begin{figure}[t]
\begin{center}
\includegraphics[width=85mm]{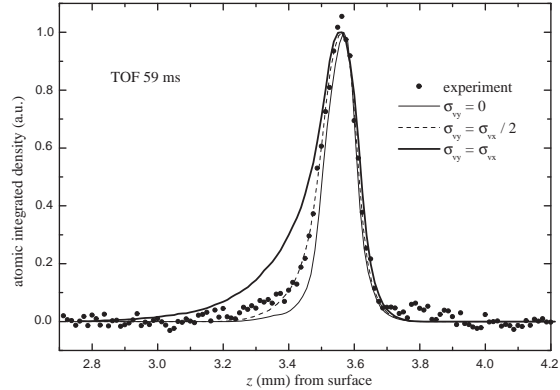}
\end{center}
\caption{Atomic density profiles integrated along $y$ and averaged 
along $x$, after 59~ms total time of flight, \textit{i.e.} 32~ms 
after reflection. Closed circles: normalized experimental data. 
Lines: result of numerical calculation with a starting height 3.59~mm 
above the mirror, $N_0/N = 0.4$, $V_y = V_z = V_{\perp} = 
5.96\,{\rm mm/s}$, $V_x = 0.89\,{\rm mm/s}$, $R_x = 90~\mu$m, $T = 
285$~nK, $v_x = -30.67\,{\rm mm/s}$, $v_z = 0.3\,{\rm mm/s}$ and 
$\sigma_{v_x} = 39\,{\rm mm/s}$, values deduced from the experimental 
pictures. Thin line: totally anisotropic scattering ($\sigma_{v_y} = 
0$); dashed line: anisotropic scattering with $\sigma_{v_y} = 
\sigma_{v_x}/2$; bold line: isotropic scattering ($\sigma_{v_y} = 
\sigma_{v_x}$). All curves are normalized to a maximum value of 
unity.}
\label{profiles}
\end{figure}

The qualitative agreement between the experimental and simulated 
pictures is very good as can be seen on figure~\ref{simul}. To
be more quantitative for the possible values of the 
velocity spread $\sigma_{v_{y}}$, we analyze the central part of the 
cloud. For 
each time of flight, a region of size 0.8~mm $\times$ 1.5~mm along 
$x$ and $z$ respectively, centred on the maximum density of the 
cloud and identical for experimental and simulated pictures, is 
isolated and an integration of the signal is performed along $x$. We 
are left with a cut of the cloud along $z$, averaged over 0.8~mm 
along 
$x$. The experimental profile is compared to the simulated one, for 
different choices of $\sigma_{v_{y}}$ after the bounce. 
Results are shown on figure~\ref{profiles} for a time of flight 59~ms.

The experimental data clearly exclude an isotropic diffuse 
reflection (figure~\ref{profiles}, bold line). They also are 
different from the pure one dimensional 
scattering case (thin line): what fits best of all is a model 
intermediate between these two extremes, \textit{i.e.} the scattering 
is only half as strong along $y$ compared to $x$. The atom mirror
thus has an angular reflection characteristic that is elongated in 
the direction parallel to the (real part of the) wave vector of the 
evanescent wave. Spontaneous emission plays only a minor role for
our parameters, but we found that the agreement with the experimental
density profiles is improved by taking it into account, in particular
on the lower left wing of the peak.

\subsection{Mirror corrugation}

\begin{figure}[t]
\begin{center}
\includegraphics[width=50mm]{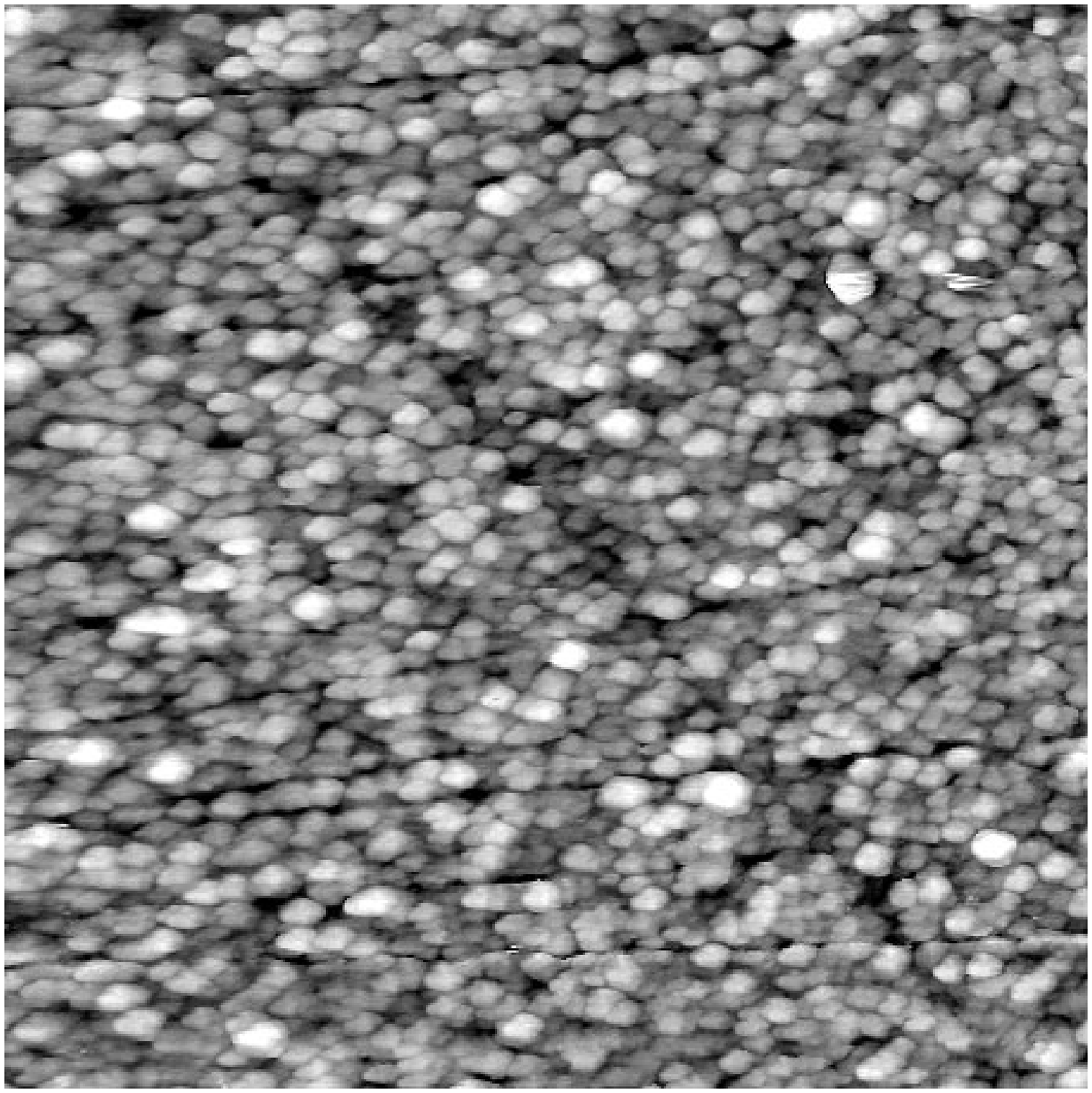}%
\hspace*{05mm}
\raisebox{-07mm}{
\includegraphics[width=85mm]{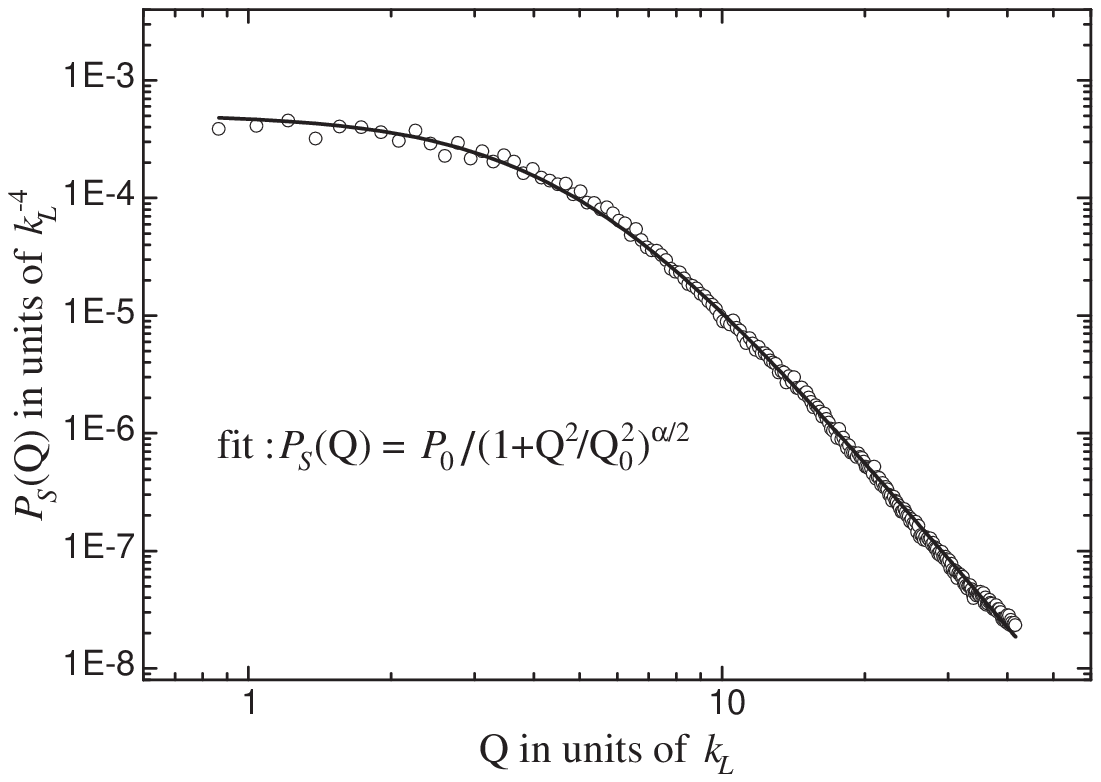}}%
\end{center}
\caption{(left) Typical AFM picture of the prism surface. The dimensions are 
$4.5\,\mu{\rm m}\times 4.5\,\mu{\rm m}$. The grains are the top facets of
pillar-like structures characteristic of an epitaxed TiO$_2$
surface.}
\label{AFMpicture}
\end{figure}

For a theoretical prediction of the anisotropic mirror reflection, we 
use the 
theory
of Ref.\cite{Henkel97a} where the diffuse scattering is attributed to
the interference between the evanescent wave and light diffusely 
scattered
from the rough glass surface. Within this theory, one can compute the
width of the momentum distribution of the reflected atoms provided the
power spectrum of the surface roughness is known. 
This power spectrum is a quantitative measure of the surface quality
and has been measured with an atomic force microscope (AFM). A typical
$4.5 \times 4.5 \mu$m$^2$ portion of the surface of the coated prism 
is shown in figure~\ref{AFMpicture}. One sees 
the top face of
pillar-like 
structures which are typical for epitaxially grown TiO$_2$ on a 
substrate. The AFM data yield a surface 
roughness $\sigma = 3.34$~nm (the rms spread of the measured surface
profile). A Fourier transform of the AFM image gives access to the 
power 
spectrum $P_S( {\bf Q} )$. (We use capitalized boldface letters for 
two-dimensional vectors in the mirror plane.)
It is found to be isotropic (a function of $Q$ only) and well fitted 
in 
the wave vector range $1\,k_L \ldots 13\,k_L$ by a power law with a 
low-frequency cut-off (see figure~\ref{Psfit})
\begin{equation}
P_S( {\bf Q} ) = \frac{P_0}{\left(1 + 
Q^2/ Q_0^2 \right)^{\alpha/2}}
\label{eq:PS-model}
\end{equation}
\begin{figure}[t]
\caption{(right) Open circles: Power spectrum $P_S(Q)$ deduced from the AFM 
measurement. Solid line: fit of the data by the function given at 
Eq.(\ref{eq:PS-model}).}
\label{Psfit}
\end{figure}%
The fit gives access to the parameters 
$\alpha=4.8$, $P_0 = 5.3 \times 10^{-4}~k_L^{-4}$ and $Q_0 = 
4.94~k_L$. In terms of this power spectrum, the rms 
surface roughness $\sigma$ is given by 
\begin{equation}
\sigma^2 = \int\!\frac{{\rm d}^2Q}{(2 \pi)^2} \, P_S( {\bf Q} ),
\end{equation}
and the fitted parameters yield $\sigma = 3.36$~nm, in excellent 
agreement with the value directly deduced from the rms spread of the 
AFM data.

\subsection{Diffuse reflection theory and comparison to the data}

We now show that the diffuse reflection we observe can be understood 
within the theory of Ref.\cite{Henkel97a}. 
We note first that the cloud is so dilute at the bounce that a
single-atom picture is sufficient to capture the physics
\cite{Bongs99}.  
For a fixed incident
momentum ${\bf p}_{\rm inc} = {\bf P} - {\bf e}_{z} m v_{i}$ near 
normal 
incidence,
the reflected wave function can be written in the form of a plane 
wave with a randomly modulated phase front:
\begin{equation}
    \psi_{\rm refl}( {\bf r} ) =  N \exp {\rm i} \left[ {\bf p}_{\rm 
    spec} \cdot{\bf r} + \delta\phi( {\bf R} ) \right],
    \label{eq:refl-psi}
\end{equation}
where $N$ is a normalization factor
and ${\bf p}_{\rm spec} = {\bf P} + {\bf e}_{z} m v_{i}$.
(The dependence on the angle of incidence
is actually negligible for our parameters~\cite{Henkel97a}.)
The phase $\delta\phi( {\bf R} )$ 
depends on the `impact position' ${\bf R}$ on the mirror, 
\textit{i.e.,} the projection of {\bf r} onto the mirror plane. We 
perform 
an ensemble average over the realizations of the rough surface and 
compute the atomic momentum distribution $P_A( {\bf P} + \hbar {\bf 
Q} )$ 
from the (spatial) Fourier transform of the `atomic coherence
function' (Sec.~6 of Ref.\cite{Henkel97a})
\begin{eqnarray}
\langle
    \psi^*_{\rm refl}( {\bf r} ) \psi_{\rm refl}( {\bf r}' )
    \rangle
    &=& 
N^2     \exp {\rm i} \left[ 
{\bf p}_{\rm spec}\cdot({\bf r}' - {\bf r}) \right]
\label{eq:atomic-coh-func}
\\
&&
\times
\exp\left[ - \frac12 \langle \left( \delta\phi( {\bf R} ) 
- \delta\phi( {\bf R}' ) \right)^2 \rangle \right],
\nonumber
\end{eqnarray}
(We take $\langle \delta\phi( {\bf R} ) \rangle = 0$, 
assuming the roughness to be statistically homogeneous.)
The variance of the phase shift can be found from the following 
formula (Eqs.(6.15) and (5.16) of Ref.\cite{Henkel97a})
\begin{equation}
    \langle \delta\phi( {\bf R} ) \delta\phi( {\bf R}' ) \rangle
=    \int\!\frac{{\rm d}^2Q}{(2 \pi)^2} \, 
P_S( {\bf Q} ) |B_{\rm at}( {\bf Q} )|^2
{\rm e}^{ {\rm i} {\bf Q} \cdot ( {\bf R} - {\bf R}' ) },
    \label{eq:phase-corrs}
\end{equation}
where $B_{\rm at}( {\bf Q} )$ is the ``atomic response function'' 
given
in Eq.(5.15) of Ref.\cite{Henkel97a}.
For the parameters of our experiment, we find that the phase shift 
has a variance $\langle \delta\phi^2( {\bf R} ) \rangle = 16.5$ large 
compared to unity. 
In this regime, Ref.\cite{Henkel97a} has shown that the reflected 
atomic 
velocity distribution approaches a Gaussian shape whose width along 
the $x$-direction, for example, is given by
\begin{equation}
\frac{ \sigma_{v_x}^2 }{ v_{\rm rec}^2 } =
\frac{1}{k_L^2}
\int\!\frac{{\rm d}^2Q}{(2 \pi)^2} \, Q_x^2 P_S( {\bf Q} ) 
|B_{\rm at}( {\bf Q} )|^2
.
\label{eq:dQx-theory}
\end{equation}
This expression gives the additional broadening of the incident 
velocity distribution due to the diffuse mirror reflection.
We perform the integration of Eq.(\ref{eq:dQx-theory}) 
numerically, with the roughness power spectrum determined previously 
from the AFM images
(Eq.(\ref{eq:PS-model})). For simplicity, we
calculate the response function $B_{\rm at}( {\bf Q} )$ using
scalar light scattering from the topmost interface only, ignoring 
the actual layered structure. We believe that this approximation
is sufficient, at least for describing the scattering in the 
$x$-direction: as shown in Ref~\cite{Henkel97a}, the
atom does not change its magnetic sublevel if it scatters in this
direction and if the evanescent wave is linearly polarized. These 
conditions are met here so that both atom and light can be 
described by scalar wave fields.

Within the theoretical model outlined above, 
the velocity spread along the propagation direction of the 
evanescent wave is found to be $\sigma_{v_x} = 6.76~v_{\rm 
rec}$. This value is in very good agreement with the experimental value 
$6.6\pm0.2~v_{\rm rec}$. This is a very satisfying result because
the theory only contains, 
within the approximations we made, parameters that are based 
on independent measurements. We believe that this is the first 
quantitative demonstration of evanescent wave scattering in the 
diffuse regime.

\subsection{Discussion of the anisotropy}

We also compute the anisotropy of the reflected 
atoms and find a ratio $\sigma_{v_x} / \sigma_{v_y} = 2.6$, in 
good agreement with the value ($2 \pm 0.5$) extracted from the
experimental data. As discussed in Ref.\cite{Henkel97a},
this anisotropy arises from
the fact that diffuse reflection occurs predominantly by Bragg
transitions where a photon is absorbed from the evanescent wave
(with wave vector $k_x = k_L n_{\rm TiO_2} \sin\theta_i$) 
and another photon is emitted into a diffusely scattered mode that 
emerges at grazing incidence into the vacuum half-space (or the inverse
process). If these scattered modes are distributed isotropically 
in the mirror plane on a circle of radius $r_{\rm sc} k_{L}$,  
the ratio of the rms spreads would be 
$\sigma_{v_{x}} / \sigma_{v_{y}} = 
(2 (n_{\rm TiO_2} \sin(\theta_i) / r_{\rm sc})^2 + 1)^{1/2}$. 
Taking $r_{\rm sc} = 1$, which 
corresponds to scattered modes emerging at grazing incidence,
we again find an anisotropy ratio of $\approx 2.5$. This agreement is 
not very surprising since the rough surface has a power spectrum much
broader than the photon wavenumber $k_{L}$ (Fig.\ref{Psfit}). 
Within this simple calculation, however, we can also get a quick 
estimate of the impact of the dielectric coating. The choice
$r_{\rm sc} = n_{\rm TiO_2} \sin(\theta_i)$ corresponds to resonant 
scattering into waveguide modes in the TiO$_{2}$ layer and 
leads to a ratio $\sigma_{v_{x}} / \sigma_{v_{y}} = \sqrt{3}$ which
cannot be excluded experimentally.

\section{Conclusion}

In conclusion, we have observed the diffuse reflection of an ultracold
atomic beam from an evanescent wave. The wave propagates on the rough 
surface of a dielectric prism, and light scattering leads to an atom 
mirror showing a significantly nonspecular reflection. The angular 
broadening of the reflected atoms, as well as their anisotropic 
angular distribution in the mirror plane, are in good agreement with 
a theory developed by Henkel \textit{et al.} \cite{Henkel97a}.
It is remarkable that 
this agreement does not imply any free parameters since we 
independently
measured the spectrum 
of the surface roughness with an AFM.
In our experiment, using a BEC has mainly practical advantages. 
Indeed, as we mentioned above, everything can be understood within a 
single-atom picture, and after diffuse scattering, spatial coherence 
is seriously reduced, as is discussed in Ref.\cite{Henkel97a} 
and investigated in Ref.\cite{Esteve04b}. Nevertheless, the BEC 
provides crucial advantages because we achieve a very clean situation.
Apart from a very low velocity spread
$V_x \ll \sigma_{v_{x}}$, a BEC has a negligible size when
impacting the evanescent wave surface. This removes the need to take
into account the mirror curvature due to the gaussian spot profile;
the contribution of the initial size to the cloud
width after reflection is negligible; and the losses given the finite
size of the mirror (the waist of the reflected laser beam) are 
minimal. In fact, with a freely falling, ultracold, but thermal gas,
the finite mirror size would lead to strongly reduced signal.

\ack
\small We gratefully acknowledge support by the R\'egion 
Ile-de-France (contract number E1213) and by the European Community 
through the Research Training Network ``FASTNet'' under contract No.\ 
HPRN-CT-2002-00304 and the Marie Curie Research Network ``Atom 
Chips'' 
under contract No.\ MRTN-CT-2003-505032. Laboratoire de Physique des 
Lasers (LPL) is Unit\'e Mixte de Recherche 7538 of Centre National de 
la 
Recherche Scientifique and Universit\'e Paris 13. 
The LPL group is a member of the Institut Francilien de Recherche 
des Atomes Froids.

%

\begin{thebibliography}{10}
\expandafter\ifx\csname url\endcsname\relax
  \def\url#1{{\tt #1}}\fi
\expandafter\ifx\csname urlprefix\endcsname\relax\def\urlprefix{URL 
}\fi
\providecommand{\eprint}[2][]{\url{#2}}

\bibitem{Reichel01b}
H\"{a}nsel W, Hommelhoff P, H\"{a}nsch T~W and Reichel J 2001 {\em 
Nature\/}
  {\bf 413} 498--501

\bibitem{Folman02}
Folman R, Kr{\"u}ger P, Schmiedmayer J, Denschlag J~H and Henkel C 
2002 {\em
  Adv. At. Mol. Opt. Phys.\/} {\bf 48} 263--356

\bibitem{Zimmermann02a}
Fort{\'a}gh J, Ott H, Kraft S and Zimmermann C 2002 {\em Phys. Rev. 
A\/} {\bf
  66} 041604(R)
\nonum
  Leanhardt A~E, Shin Y, Chikkatur A~P, Kielpinski D, Ketterle W and 
Pritchard
    D~E 2003 {\em Phys. Rev. Lett.\/} {\bf 90} 100404
    \nonum
    Schumm T, Est\`eve J, Figl C, Trebbia J~B, Aussibal C, Nguyen H,
    Mailly D, Bouchoule I, Westbrook C and Aspect A 2005 {\em Eur.
    Phys.  J. D\/} {\bf 32} 171--80

    \bibitem{Jones03}
    Jones M~P~A, Vale C~J, Sahagun D, Hall B~V and Hinds E~A 2003 
{\em Phys. Rev.
      Lett.\/} {\bf 91} 080401
\nonum
    Harber D~M, McGuirk J~M, Obrecht J~M and Cornell E~A 2003 {\em J. 
Low Temp.
      Phys.\/} {\bf 133} 229--38
\nonum
    Rekdal P~K, Scheel S, Knight P~L and Hinds E~A 2004 {\em Phys. 
Rev. A\/} {\bf
      70} 013811

\bibitem{Henkel99c}
  Henkel C and Wilkens M 1999 {\em Europhys. Lett.\/} {\bf 47} 414--20
\nonum
Henkel C, P{\"o}tting S and Wilkens M 1999 {\em Appl. Phys. B\/} {\bf 
69}
  379--87

\bibitem{Balykin88}
Balykin V~I, Letokhov V~S, Ovchinnikov Y~B and Sidorov A~I 1988 {\em 
Phys. Rev.
  Lett.\/} {\bf 60} 2137--40
\nonum
  Kasevich M~A, Weiss D~S and Chu S 1990 {\em Opt. Lett.\/} {\bf 15} 
607--609
    \nonum
  Aminoff C~G, Steane A~M, Bouyer P, Desbiolles P, Dalibard J and 
Cohen-Tannoudji
    C 1993 {\em Phys. Rev. Lett.\/} {\bf 71} 3083--6
    \nonum
  Landragin A, Courtois J~Y, Labeyrie G, Vansteenkiste N, Westbrook 
C~I and
    Aspect A 1996 {\em Phys. Rev. Lett.\/} {\bf 77} 1464--7

\bibitem{Dalibard96}
Christ M, Scholz A, Schiffer M, Deutschmann R and Ertmer W 1994 {\em 
Opt.
  Commun.\/} {\bf 107} 211--7
\nonum
Brouri R, Asimov R, Gorlicki M, Feron S, Reinhardt J, Lorent V and 
Haberland H
  1996 {\em Opt. Commun.\/} {\bf 124} 448--51
  \nonum
  %
Szriftgiser P, Gu\'{e}ry-Odelin D, Arndt M and Dalibard J 1996 {\em 
Phys. Rev.
  Lett.\/} {\bf 77} 4--7
  \nonum
Cognet L, Savalli V, Horvath G~Z~K, Holleville D, Marani R, Westbrook 
N,
  Westbrook C~I and Aspect A 1998 {\em Phys. Rev. Lett.\/} {\bf 81} 
5044--5047


\bibitem{Ovchinnikov91}
Ovchinnikov Y~B, Shul'ga S~V and Balykin V~I 1991 {\em J. Phys.~B: 
Atom. Mol.
  Opt. Phys.\/} {\bf 24} 3173--8
\nonum
  Gauck H, Hartl M, Schneble D, Schnitzler H, Pfau T and Mlynek J 
1998 {\em Phys.
    Rev. Lett.\/} {\bf 81} 5298--301
    \nonum
  Hammes M, Rychtarik D, Engeser B, N\"agerl H~C and Grimm R 2003 
{\em Phys. Rev.
    Lett.\/} {\bf 90} 173001 


\bibitem{Prentiss00}
Dekker N~H, Lee C~S, Lorent V, Thywissen J~H, Smith S~P, Drndi{\'c} M,
  Westervelt R~M and Prentiss M 2000 {\em Phys. Rev. Lett.\/} {\bf 84}
  1124--7

\bibitem{Grimm04a}
Rychtarik D, Engeser B, N\"agerl H~C and Grimm R 2004 {\em Phys. Rev. 
Lett.\/}
  {\bf 92} 173003 

\bibitem{Pasquini04}
Pasquini T~A, Shin Y~I, Sanner C, Saba M, Schirotzek A, Pritchard D~E 
and
  Ketterle W 2004 {\em Phys. Rev. Lett.\/} {\bf 93} 223201 
\nonum
Pasquini T~A, Saba M, Jo G, Shin Y, Ketterle K, Pritchard D~E,
Savas T~A and Mulders N 2006,
``Low velocity quantum reflection of Bose-Einstein condensates'' 
\textit{preprint} cond-mat/0603463.

\bibitem{Cornell04a}
McGuirk J~M, Harber D~M, Obrecht J~M and Cornell E~A 2004 {\em Phys. 
Rev. A\/}
  {\bf 69} 062905 

\bibitem{Vuletic04}
Lin Y~J, Teper I, Chin C and Vuleti{\'c} V 2004 {\em Phys. Rev. 
Lett.\/} {\bf
  92} 050404
  \nonum
Harber D~M, Obrecht J~M, McGuirk J~M and Cornell E~A 2005 {\em Phys. 
Rev. A\/}
  {\bf 72} 033610
\nonum
Obrecht J~M, Wild R~J, Antezza M, Pitaevskii L~P, Stringari S and 
    Cornell E~A 2006,
  ``Measurement of the Temperature Dependence of the Casimir-Polder Force''
\textit{preprint} physics/0608074.
  
\bibitem{Landragin96b}
Landragin A, Labeyrie G, Henkel C, Kaiser R, Vansteenkiste N, 
Westbrook C~I and
  Aspect A 1996 {\em Opt. Lett.\/} {\bf 21} 1581--3
\nonum
In these experiments, some indications for anisotropic scattering 
after 
reflection of thermal atoms on an eva\-nescent wave mirror were 
observed (C.~Westbrook and A.~Landragin, private communication).

\bibitem{HindsHughes}
Hinds E~A and Hughes I~G 1999,
{\em J. Phys. D: Appl. Phys.\/} {\bf 32} R119--46

\bibitem{Arnold02}
Arnold A~S, MacCormick C and Boshier M~G 2002,
{\em Phys. Rev. A\/} {\bf 65} 031601

\bibitem{Savalli2002}
Savalli V, Stevens D, Est\`eve J, Featonby P D, Josse V, Westbrook N, 
Westbrook C~I and
  Aspect A 2002 {\em Phys. Rev. Lett.\/} {\bf 88} 250404 

\bibitem{Henkel97a}
Henkel C, M{\o}lmer K, Kaiser R, Vansteenkiste N, Westbrook C~I and 
Aspect A
  1997 {\em Phys. Rev. A\/} {\bf 55} 1160--78

\bibitem{Perrin05a}
Perrin H, Colombe Y, Mercier B, Lorent V and Henkel C
2005 {\em J. Phys.: Conf. Ser.} {\bf 19} 151--7; 
doi:10.1088/1742-6596/19/1/025, \textit{preprint} 
quant-ph/0509200 

\bibitem{Kaiser94}
Kaiser R, L\'evy Y, Vansteenkiste N, Aspect A, Seifert W, Leipold D 
and Mlynek
  J 1994 {\em Opt. Commun.\/} {\bf 104} 234

\bibitem{Perrin03}
Colombe Y, Kadio D, Olshanii M, Mercier B, Lorent V and Perrin H 2003 
{\em J.
  Opt. B: Quantum Semiclass. Opt.\/} {\bf 5} S155--63

\bibitem{Castin96}
Castin Y and Dum R 1996 {\em Phys. Rev. Lett.\/} {\bf 77} 5315--9
\nonum
Kagan Y, Surkov E~L and Shlyapnikov G~V 1996 {\em Phys. Rev. A\/} 
{\bf 54}
  R1753--6
  
\bibitem{note_spont} 
This value is deduced from an integration of the
number of scattered photons along the mean classical atomic
trajectory, calculated from the known evanescent wave parameters. We
neglect the variation of the spontaneous emission rate at the vicinity
of the surface. This assumption is reasonable as the classical turning
point is rather far from the surface ($k_L z_0=1.33$). See Henkel C
and Courtois J-Y 1998 {\em Eur. Phys. J. D\/} {\bf 3} 129--153.

\bibitem{Bongs99}
Bongs K, Burger S, Birkl G, Sengstock K, Ertmer W, Rzazewski K, 
Sanpera A, and Lewenstein M 1999 
{\em Phys. Rev. Lett.\/} {\bf 83} 3577; Busch T, private communication

\bibitem{Esteve04b}
Est{\`e}ve J, Stevens D, Aussibal C, Westbrook N, Aspect A and 
Westbrook C~I
  2004 {\em Eur. Phys. J. D\/} {\bf 31} 487--91

\end{thebibliography}

\bigskip\

\providecommand{\newblock}{}

\end{document}